\def\isarxiv{1} 

\ifdefined\isarxiv
\documentclass[11pt]{article}

\usepackage[numbers]{natbib}

\else
\documentclass{article}
\usepackage{neurips_2022}
\fi

\usepackage{amsmath}
\usepackage{amsthm}
\usepackage{amssymb}
\usepackage{algorithm}
\usepackage{subfig}
\usepackage{algpseudocode}
\usepackage{graphicx}
\usepackage{grffile}
\usepackage{wrapfig,epsfig}
\usepackage{url}
\usepackage{xcolor}
\usepackage{epstopdf}

\usepackage{bbm}
\usepackage{dsfont}

\allowdisplaybreaks

\ifdefined\isarxiv

\usepackage{tikz}
\usepackage{hyperref}  
\hypersetup{colorlinks=true,citecolor=blue,linkcolor=blue} 
\usetikzlibrary{arrows}
\usepackage[margin=1in]{geometry}

\else

\usepackage{microtype}
\usepackage{hyperref}
\definecolor{mydarkblue}{rgb}{0,0.08,0.45}
\hypersetup{colorlinks=true, citecolor=mydarkblue,linkcolor=mydarkblue}

\fi

\newtheorem{theorem}{Theorem}[section]
\newtheorem{lemma}[theorem]{Lemma}
\newtheorem{definition}[theorem]{Definition}

\newtheorem{remark}[theorem]{Remark}
\newtheorem{claim}[theorem]{Claim}

\newcommand{\wh}{\widehat}
\newcommand{\wt}{\widetilde}

\newcommand{\R}{\mathbb{R}}

\renewcommand{\hat}{\wh}

\DeclareMathOperator{\nnz}{nnz}

\DeclareMathOperator{\vect}{vec}

\DeclareMathOperator{\diag}{diag}

\makeatletter
\newcommand*{\RN}[1]{\expandafter\@slowromancap\romannumeral #1@}
\makeatother

\usepackage{lineno}

\begin{document}

\ifdefined\isarxiv

\date{}

\title{Faster Sinkhorn’s Algorithm with Small Treewidth}
\author{
Zhao Song\thanks{\texttt{zsong@adobe.com}. Adobe Research.}
\and 
Tianyi Zhou\thanks{\texttt{t8zhou@ucsd.edu}. UCSD.}
}

\else

\title{Intern Project} 
\maketitle 
\fi

\ifdefined\isarxiv
\begin{titlepage}
  \maketitle
  \begin{abstract}

Computing optimal transport (OT) distances such as the earth mover’s distance is a fundamental problem in machine learning, statistics, and computer vision. In this paper, we study the problem of approximating the general OT distance between two discrete distributions of size $n$. Given the cost matrix $C=AA^\top$ where $A \in \R^{n \times d}$, we proposed a faster Sinkhorn's Algorithm to approximate the OT distance when matrix $A$ has treewidth $\tau$. To approximate the OT distance, our algorithm improves the state-of-the-art results [Dvurechensky, Gasnikov, and Kroshnin ICML 2018] from $\wt{O}(\epsilon^{-2} n^2)$ time to $\wt{O}(\epsilon^{-2} n \tau)$ time.

  \end{abstract}
  \thispagestyle{empty}
\end{titlepage}

{\hypersetup{linkcolor=black}
\tableofcontents
}
\newpage

\else

\begin{abstract}

Computing optimal transport (OT) distances such as the earth mover’s distance is a fundamental problem in machine learning, statistics, and computer vision. In this paper, we study the problem of approximating the general OT distance between two discrete distributions of size $n$. Given the cost matrix $C=AA^\top$ where $A \in \R^{n \times d}$, we proposed a faster Sinkhorn's Algorithm to approximate the OT distance when matrix $A$ has treewidth $\tau$. To approximate the OT distance, our algorithm improves the state-of-the-art results [Dvurechensky, Gasnikov, and Kroshnin ICML 2018] from $\wt{O}(\epsilon^{-2} n^2)$ time to $\wt{O}(\epsilon^{-2} n \tau)$ time.

\end{abstract}

\fi

\section{Introduction}

Optimal transport is a mathematical theory that deals with the problem of finding the most efficient way to transport goods or materials from one place to another. The goal is to minimize the cost of transportation, which is usually measured in terms of the distance traveled or the amount of resources used. Many problems in computational sciences require to use optimal transport to compare between probability measures or histograms, including Wasserstein or earth mover’s distance \cite{wpr85,rtg00,v09}. Optimal transoport has a wide range of application, such as bag-of-words for natural language processing \cite{kskw15}, multi-label classification \cite{fzm+15}, unsupervised learning \cite{acb17,bgkl17}, semi-supervised learning \cite{srgb14}, statistics \cite{ess17,pz16}, and other application \cite{kpt+17}.
In particular due to its applications in image processing, it has recently become crucial to have efficient ways of computing, or approximating, the optimal transport or the Wasserstein distances between two measures. 

There is a long line of research on OT problem. \cite{c13} apply Sinkhorn’s algorithm to the entropy-regularized OT optimization problem. As it was recently shown in \cite{anr17}, this approach allows to find an $\epsilon$-approximation for an OT distance in $\wt{O}(\epsilon^{-3} n^2)$ time. In terms of the dependence on $n$, this result improves on the complexity $\wt{O}(n^3)$ achieved by the network simplex method or interior point methods \cite{pw09}, applied directly to the OT optimization problem, which is a linear program \cite{j42}. The cubic dependence on $\epsilon$ prevents approximating OT distances with good accuracy. Then, in \cite{dgk18}, they proposed an algorithm with the complexity bound $\wt{O}(\epsilon^{-2} n^2)$ based on the Sinkhorn's algorithm. 

The treewidth of a matrix is a measure of the complexity of its structure and plays a crucial role in the design and analysis of algorithms for manipulating and processing matrices. In particular, the treewidth of a matrix can be used to determine the efficiency of algorithms that rely on tree decompositions, such as dynamic programming and divide-and-conquer techniques. In the small treewidth setting, algorithms for matrix manipulation and processing can often achieve near-linear running time, making them highly efficient and scalable. This has important implications for a wide range of applications, including interior point methods \cite{gs22,dly21}, computing John ellipsoid \cite{syyz22}, streaming algorithm \cite{lszzz22}. 
 Treewidth is also important in graph structure theory, particularly in the study of graph minors by Robertson and Seymour \cite{rs10}. Many results \cite{b94} have shown that NP-hard problems can be solved in polynomial time on classes of graphs with bounded treewidth.

The best previous work to solve this problem requires $n^2$. It is natural to ask a question 
\begin{center}
{\it Is that possible to solve in $o(n^2)$ under some mild assumption, e.g. tree width}
\end{center}
In this paper, we provide a positive answer for this question. The comparison between our results and previous work's is shown in Table \ref{tab:history}.

\begin{table*}[!ht]
    \centering
    \begin{tabular}{|l|l|l|l|l|} \hline 
        {\bf References} & {\bf Method} & {\bf Time Complexity}  \\ \hline 
        \cite{pw09} & Network Simplex Method &  $n^3$  \\ \hline
       \cite{anr17} & Sinkhorn’s algorithm & $\epsilon^{-3} n^2$ \\ \hline 
       \cite{dgk18} & Sinkhorn’s algorithm &$\epsilon^{-2} n^2$ \\ \hline
       Theorem \ref{thm:runningtime_ot} & Sinkhorn’s algorithm & $\epsilon^{-2} n \tau $ \\ \hline
    \end{tabular}
    \caption{Given the cost matrix $C = AA^\top \in \R^{n \times n}$, let $\tau$ denote the treewdith of matrix $A$. Let $\epsilon$ denote the accuracy parameter. Since $\tau \leq n$, our algorithm (Theorem \ref{thm:runningtime_ot}, Algorithm \ref{alg:ot_dis}) is  always better than \cite{dgk18}.}
    \label{tab:history}
\end{table*}

\subsection{Our Result}
We formally state our main theorem

\begin{theorem}\label{thm:informal_runningtime_ot}
    Given the cost matrix $C=AA^\top$ where $A$ has treewidth $\tau$, we can find the transport plan for the $\epsilon$-approximation of the optimal transport distance in  
    \begin{align*}
        O( \epsilon^{-2} n\tau \|C\|_\infty^2 \ln n  )
    \end{align*}
    time.
\end{theorem}

Comparing with \cite{dgk18}, that solves the problem in $O(\epsilon^{-2} n^2 \|C\|_\infty^2 \ln n)$, we proposed an algorithm that constructing matrix using its implicit form. By leveraging the property of low treewidth, our running time has no dependence on $n^2$.
\subsection{Related Work}

\paragraph{OT Problems}
OT distances, which is also called Earth Mover’s Distances \cite{rtg00}, are progressively being adopted as an effective tool in a wide range of situations, from computer graphics \cite{bpc16} to supervised learning \cite{fzm+15}, unsupervised density fitting \cite{bbr06} and generative model learning (\cite{mmkc16,acb17,szrm18,gpc18,sbrl18}).
 There is a long line of work on reducing the time complexity for solving OT.
 In \cite{acb17}, they proved that, for regularized OT, the near-linear time complexity can be achieved by both Sinkhorn and Greenkhorn algorithm. 
 They demonstrated that both algorithms have a complexity of $\wt{O}(\epsilon^{-3} n^2)$, where $n$ represents the number of atoms (or the dimension) of the probability measure being considered and $\epsilon$ is the desired level of tolerance.
 In \cite{dgk18}, the complexity of the Sinkhorn algorithm was improved to $\wt{O}(\epsilon^{-3} n^2)$. Additionally, an adaptive primal-dual accelerated gradient descent (APDAGD) algorithm was introduced, that was shown to have a complexity of $\wt{O}(\min \{ \epsilon^{-1} n^{9/4}, \epsilon^{-2} n^2 \})$.
 With a carefully designed Newton-type algorithm, \cite{alow17,cmtv17} solve the OT problem by making use of a connection to matrix-scaling problems. \cite{bjks18,q18} gave a complexity bound of $\wt{O}(\epsilon^{-1} n^2)$ for Newton-type algorithms.

\paragraph{Treewidth Problems}
Treewidth is a concept from structural graph theory that has been studied in relation to fixed parameter tractable algorithms in various fields, including combinatorics, integer-linear programming, and numerical analysis. 
\cite{fls+18} shows several problems can be reduced to matrix factorizations efficiently, including computing determinant, computing rank, and finding maximum matching, and this leads to $O(\tau^{O(1)} \cdot n)$ time algorithms where $\tau$ is the width of the given tree decomposition of the graph.
\cite{b94} shows a number of NP-hard problems such as \textsc{Independent Set, Hamiltonian Circuit, Steiner Tree, and Travelling Salesman} can be solved with run-times that depend only linearly on the problem size and exponentially on treewidth as the result of dynamic programming. By leveraging the small treewidth setting, \cite{dly21} proposed an algorithm that solves the linear program problem  with run-time nearly matching the fastest run-time for solving the sub-problem $Ax = b$. 
\cite{lszzz22} proposed a space-efficient interior point method (IPM) in the streaming model. For the linear programs with treewidth $\tau$, they solve them in $\wt{O}(n\tau)$ space, where $n$ is the number of dimension for the feature space.
\cite{syyz22} shows that, when the constraints matrix has treewidth $\tau$, the John Ellipsoid problem can be solved in $O(n \tau^2)$ time. 
The small treewidth setting is also applied to solve the semidefinite program. In \cite{gs22}, they give the first SDP solver that runs in time in linear in number of variables under this setting.

\subsection{Technique Overview}

\paragraph{Analysis}
We first provide the bounds for $u_k,v_k$ and an optimal solution $(u^*,v^*)$ for Eq.~\eqref{eq:min_pro}.
Then, we introduce the convex function of $(\hat{u},\hat{v})$ as the following:
\begin{align*}
    \langle \mathbf{1}_n,B(\hat{u},\hat{v}) \mathbf{1}_n \rangle - \langle \hat{u},B(u_k,v_k) \mathbf{1}_n \rangle - \langle \hat{v},B(u_k,v_k)^\top \mathbf{1}_n \rangle.
\end{align*}
The gradient for the above function vanishes when $(u^*,v^*)=(u_k,v_k)$, so the point $(u_k,v_k)$ is the minimizer of this function.

Therefore, we can show that 
\begin{align*}
    \wt{\psi}(u_k,v_k)  \leq & ~ \langle u_k-u_*, B_k \mathbf{1}_n -r \rangle + \langle v_k - v_*, B_k^\top \mathbf{1}_n-c\rangle
\end{align*}
Then, for each iteration of the algorithm, we upper bound the r.h.s. and get
\begin{align*}
        \wt{\psi} (u_k,v_k) \leq R \cdot (\|B_k \mathbf{1}_n - r\|_1 + \| B_k^\top \mathbf{1}_n - c\|_1).
\end{align*}
where the inequality follows from the bounds for the iterates $u_k,v_k$ and an optimal solution $(u^*,v^*)$.

Next, by using this upper bound for $\wt{\psi}$ and Lemma \ref{lem:pinsker} we have:
\begin{align*}
    &~ \wt{\psi}(u_k, v_k)-\wt{\psi}(u_{k+1}, v_{k+1})\notag\\
    \geq &~ \max \{ \frac{\wt{\psi}(u_k,v_k)^2}{2R^2},\frac{ \epsilon_0^2 }{2} \},
\end{align*}

By using induction, we prove the potential function $\wt{\psi}$ is also upper bounded by $\frac{2R^2}{k+\ell-1}$, where $\ell=\frac{2 R^2}{\wt{\psi} (u_1, v_1 )}$. 
Finally, by using the switching strategy, we provide the upper bound of the total number of iterations $k$ for the Sinkhorn's algorithm as the following
\begin{align*}
    k \leq 2+\frac{4R}{\epsilon_0}.
\end{align*}

\paragraph{Running time}
Given the cost matrix $C=MM^\top$ where $M \in \R^{n \times d}$ has treewidth $\tau$, we leverage the fact that it admits a succinct Cholesky factorization and $\nnz(C) = O(n \tau)$. 

For each iteration in Sinkhorn's algorithm (Algorithm \ref{alg:sinkhorn}), we have to compute $B(u,v)=\diag(e^u) K \diag(e^v)$ where $K_{i,j} := \exp( -C_{i,j} /\gamma )$. In fact, writing down $K$ explicitly requires $O(n^2)$. To bypass this issue, we first write $K$ in implicit form $K_{i,j} := A_{i,j} - D_{i,j}$, where $ A_{i,j} = e^{-C_{i,j} /\gamma} - 1$ and $D_{i,j} =  1$, so that matrix $A$ is as sparse as matrix $C$. Also, we represent matrix $D$ by $ww^\top$, where $w = \mathbf{1}_n$. Leveraging the fact that $\nnz(A) = O(n \tau)$ and matrix $D$ is a rank-1 matrix. We improve the per iteration running time for Sinkhorn algorithm from $O(n^2)$ to $O(n \tau)$. 

For the rounding algorithm (Algorithm \ref{alg:round}) of the transport plan $B$, we also write down the transport plan in an implicit fashion and do the computation in $O(n \tau)$ time. Note that we \emph{never} write down $B,B_0,B_1$ and output $G$ explicitly. When computing $B \mathbf{1}_n$, we leverage the implicit form of $B$ and do the computation as following:
\begin{align*}
    \diag(e^{u_k}) A \mathbf{1}_n \diag(e^{v_k})  +\diag(e^{u_k}) (w w^\top) \mathbf{1}_n \diag(e^{v_k}).
\end{align*}
As $\nnz(A) = O(n \tau)$, computing $A \mathbf{1}_n$ takes $O(n \tau)$ time. Similarly, when computing $X B$, where $X$ is a diagonal matrix, we leverage the implicit form of $B$ and do the computation as following:
\begin{align*}
    \diag(e^{u_k}) A X \diag(e^{v_k})  +\diag(e^{u_k}) (w w^\top) X \diag(e^{v_k}).
\end{align*} As $\nnz(A) = O(n \tau)$, computing $A X$ takes $O(n \tau)$ time and the $AX$ is also $O(n\tau)$ sparse.

Finally, we note that with $\wt{O}(\epsilon^{-2} n \tau)$ time we approximate the transport plan for the OT distance problem.

{\bf Roadmap.}

We first introduce all required preliminary in Section \ref{sec:pre}. Then, we provide the analysis for the Sinkhorn's algorithm in Section \ref{sec:sink}. In Section \ref{sec:ot_treewidth}, we provide the faster Sinkhorn's algorithm with small treewidth setting and apply our faster Sinkhorn's Algorithm to solve the OT distance. 

\section{Preliminary}\label{sec:pre}
For a positive integer $n$, we denote $[n]=\{1,2,\cdots,n\}$. We use ${\bf 1}_n$ denote the length-$n$ vector where all the entries that are ones. 

For a vector $a$, we denote $e^a,\ln a$ as their entry-wise exponents and natural logarithms respectively. We define $a_{k,i}$ as the $i$-th coordinate of $k$-th iteration of the $a$.

For a matrix $A \in \R^{n \times n}$, we define $\| A \|_{\infty}:= \max_{i,j \in [n]} | A_{i,j} |$. 
We define $A_{i,j}$ as the entry at $i$-th row and $j$-th coloum of matrix $A$.
We use $e^A,\ln A$ to denote their entry-wise exponents and natural logarithms respectively.
We denote by $\vect(A)$ the vector in $\R^{n^2}$,
which is obtained from $A$ by writing its columns one below another.
For two matrices $A,B$, we denote their  inner product by $\langle A,B \rangle$.
We define the $n$-dimensional simplex as $\triangle _n := \{x \in \R^n_+: \sum_{i=1}^n x_i = 1\}$. 

For a vector $x \in \R^n$, we define its $\ell_p$ norm to be $\| x \|_p:= ( \sum_{i=1}^n |x_i|^p )^{1/p}$. For two vectors $x,y$, we define the inner product $\langle x, y \rangle = \sum_{i=1}^n x_i y_i$.

The definition of entropy is given as the following:
\begin{definition}[Entropy]\label{def:entropy}
    We define the entropy $H(p)$ of vector $p$ by 
    \begin{align*}
        H(p):=\sum_{i=1}^n p_i \log (\frac{1}{p_i}).
    \end{align*} Similarly, for a matrix $P \in \R_{+}^{n \times n}$, we define the entropy $H(P)$ entrywise as 
    \begin{align*}
        \sum_{i=1}^n \sum_{j=1}^n \log \frac{1}{P_{i,j}}.
    \end{align*}
\end{definition}

\subsection{Problem Formulation}
We first introduce the definition of OT problem.
\begin{definition}
Given a matrix $C$ with small tree width (e.g. $C = A A^\top$ where $A \in \R^{n \times d}$), the optimal transport problem is defined as:
\begin{align*}
 \min_{X} & ~ \langle C, X \rangle \\
 \mathrm{~s.t.~} & ~ X \in \R_{+}^{n \times n} \\
 & ~ X {\bf 1}_n = r \\
 & ~ X^\top {\bf 1}_n = c
\end{align*}
where ${\bf 1}_n \in \R^n$ denotes a vector where every entry is $1$.
\end{definition}

Next, we give the definition of the regularized OT problem.
\begin{definition}
    Given a strongly convex regularizer $\mathcal{R}(X)$, e.g. negative entropy or squared Euclidean norm, the regularized optimal transport problem is defined as: 
\begin{align}\label{eq:reg_ot}
 \min_{X} & ~ \langle C, X \rangle + \gamma \mathcal{R}(X) \\
 \mathrm{~s.t.~} & ~ X \in \R_{+}^{n \times n} \notag\\
 & ~ X {\bf 1}_n = r \notag\\
 & ~ X^\top {\bf 1}_n = c \notag
\end{align}
where $\gamma>0$ denotes the regularization parameter. 
\end{definition}
 
The goal for this paper is to find the approximation for the transportation plan $\hat{X}$ defined as follows: 
\begin{definition}[$\epsilon$-approximation]
    The $\epsilon$-approximation for the OT distance is defined as \begin{align}\label{eq:app_ot}
     \langle C, \hat{X} \rangle \leq & ~ \min_{X}  \langle C, X \rangle + \epsilon \\
     \mathrm{~s.t.~} & ~ X \in \R_{+}^{n \times n} \notag\\
     & ~ X {\bf 1}_n = r \notag\\
     & ~ X^\top {\bf 1}_n = c \notag
    \end{align}
    where $\hat{X}$ denotes the approximation for the transportation plan.
\end{definition}

For simplicity we introduce the definition of $\mathcal{U}_{r,c} \subset \R_{+}^{n \times n}$ 
\begin{definition}\label{def:u_r_c}
    Given the OT problem $\arg\min_{X \in \mathcal{U}_{r,c}}\langle X,C \rangle $, we define
    \begin{align*}
         \mathcal{U}_{r,c} := \{X \in \R^{n \times n}_+: X \mathbf{1}_n=r, X^\top \mathbf{1}_n = c\}
    \end{align*}
    where $\mathbf{1}_n$ is the all-ones vector in $\R^n$ , $C\in \R^{n\times n}_+$ is a given cost matrix, and $r \in \R ^n ,c \in \R^n$ are given vectors with positive entries that sum to one.
\end{definition}

Next, we provide a lemma about the transport plan $X$.
\begin{lemma}[\cite{c13}]\label{lem:unique_min}
For any cost matrix $C \in \R^{n \times n}$, $\mathcal{U}_{r,c} \subset \R_{+}^{n \times n}$ and $r,c \in \triangle_n$, the minimization program 
\begin{align*}
     X_\gamma := \arg\min_{X \in \mathcal{U}_{r,c}}\langle X,C \rangle + \gamma \cdot \mathcal{R}(X) ,
\end{align*}
where $\gamma>0$ is the regularization parameter and $\mathcal{R}(X)$ is a strongly convex regularizer,
has a unique minimum at $X_\gamma \in \mathcal{U}_{r,c}$ of the form $X_\gamma = MAN$, where $A := \exp (-\frac{1}{\gamma} C)$ and $M,N \in \R_{+}^{n \times n}$ are both diagonal matrices. The matrices $(M,N)$ are unique up to a constant factor.
\end{lemma}

\subsection{Inequalities}
We introduce the H\"{o}lder's inequality as following:
\begin{lemma}[H\"{o}lder's inequality]\label{lem:holder}
    If $p>1$ and $q>1$ are such that 
    \begin{align*}
        \frac{1}{p} + \frac{1}{q} = 1
    \end{align*}
    then
    \begin{align*}
        \|ab\|_1  \leq \|a\|_p \|b\|_q.
    \end{align*}
\end{lemma}
We also provide the Pinsker inequality.
\begin{lemma}[Pinsker inequality]\label{lem:pinsker}
Let $P$ and $Q$ be two distributions defined on the universe $U$. Then,
\begin{align*}
    \mathrm{KL}(P\|Q) \geq \frac{1}{2 \ln 2} \cdot \|P-Q\|_1^2.
\end{align*}
where $\mathrm{KL}(P\|Q)$ is the $\mathrm{KL}$-divergence between $P$ and $Q$.
\end{lemma}




\subsection{Treewidth preliminaries}

We begin by introducing the definition of treewidth for a given matrix.  
\begin{definition}[Treewidth $\tau$]\label{def:treewidth} 
Given a matrix $A \in \R^{n \times d}$, we construct its graph $G = (V,E)$ as follows: The vertex set are columns $[d]$; An edge $(i,j) \in E$ if and only if there exists $k \in [n]$ such that $A_{k,i} \ne 0, A_{k,j}\ne 0$. Then, the treewidth of the matrix $A$ is the treewidth of the constructed graph.
In particular, every column of $A$ is $\tau$-sparse. 
\end{definition}

Next, we present the definition for Cholesky factorization.

\begin{definition}[Cholesky Factorization]\label{def:cholesky_dec}
Given a positive-definite matrix $P$, there exists a unique Cholesky factorization $P = LL^\top \in \R^{d \times d}$, where $L \in \R^{d \times d}$ is a lower-triangular matrix with real and positive diagonal entries. 
\end{definition}

We also provide the running time of computing the Cholesky factorization.

\begin{lemma}[\cite{gln94,dly21}]\label{lem:cholesky_time}
Given a positive definite matrix $M \in \R^{d \times d}$, we can decompose it by using Cholesky decomposition $M=LL^\top$ in time 
\begin{align*}
    \Theta(\sum_{j=1}^d | \mathcal{L}_j |^2),
\end{align*}
where $| \mathcal{L}_j|$ is the number of nonzero entries in the $j$-th column of $L$.
\end{lemma}

Then, we introduce some results based on the Cholesky factorization of a given matrix with treewidth $\tau$: 

\begin{lemma}[\cite{bghk95,d06,dly21}]\label{lem:fast_cholesky}
For any matrix $A \in \R^{n \times n}$ with treewidth $\tau$, we can compute the Cholesky factorization $A A^\top = L L^\top \in \R^{n \times n}$ in $O(n \tau^2)$ time,  where $L \in \R^{n \times n}$ is a lower-triangular matrix with real and positive entries.
$L$ satisfies the property that every column is $\tau$-sparse.
\end{lemma}

\begin{claim}[\cite{dly21,gs22,syyz22,lszzz22}]\label{cla:nnz_c}
    Given $L=MM^\top$, where $M$ has treewidth $\tau$ and $M \in \R^{m \times n}$, we have $\nnz(L) = O( n \tau)$.
\end{claim}
\begin{proof}
    We first show that $\nnz(L) = O(m)$.
    Let $M \in \R^{m \times n }$ denote the adjacency matrix of graph $G=(V,E)$, where $|E(G)| = m, |V(G)| = n$. The Laplacian matrix of graph $G$ is $L = MM^\top$ and it is also defined as $D - A$, where $D$ is the degree matrix and $A$ is the adjacency matrix of graph $G$. As $\nnz(A) = O(m), \nnz(D) = O(n)$ and $m \geq n$, we have 
    \begin{align}\label{eq:nnz_l}
        \nnz(L) = O(m)+O(n) = O(m).
    \end{align}
    Next, we show that the number of edge $m$ for graph $G$ is bounded by $O(n \tau)$. The maximal graphs with treewidth $\tau$ are the $\tau$-trees which are constructed by starting with a $(\tau+1)$-clique and iteratively adding vertices of degree $\tau$ such that its neighbours form a $\tau$-clique.  By counting the edges in the $(\tau+1)$-clique and the edges incident to the $n - \tau - 1$ vertices iteratively added to the $\tau$-tree, the total number of edges in a $\tau$-tree with $n$ vertices is 
    \begin{align}\label{eq:tau_1}
        {\tau+1 \choose 2}+ \tau(n - \tau -1) = O( n \tau).
    \end{align}
    Since any graph G with treewidth $\tau$ is a subgraph of a $\tau$-tree, we have $O( n \tau)$ is an upper bound on $|E(G)|=m$.
    By combining Eq.~\eqref{eq:nnz_l} and Eq.~\eqref{eq:tau_1}, we have $\nnz(L) = O(n \tau)$. 
    
    Hence, we complete the proof.
\end{proof}

\section{Sinkhorn's Algorithm Analysis}\label{sec:sink}

\begin{algorithm}[!ht]
\caption{Sinkhorn’s Algorithm}\label{alg:sinkhorn}
\begin{algorithmic}[1]
\Procedure{SinkhornAlgorithm}{$c,r,\epsilon_0$}  \Comment{Theorem \ref{thm:upper_k}}
\State \Comment{Accuracy $\epsilon_0$ 
}
\State $k \gets 0 $
\State $u_0 \gets 0$
\State $v_0 \gets 0$ 
\While{$\|B(u_k,v_k) \mathbf{1}_n -r\|_1 + \|B(u_k,v_k)^\top \mathbf{1}_n-c\|_1 \ge \epsilon_0$}
\If {$k \mod 2 =0$}
\State $u_{k+1} \gets u_k + \ln r - \ln (B(u_k,v_k) \mathbf{1}_n)$
\State $v_{k+1} \gets v_k$
\Else
\State $v_{k+1} \gets v_k + \ln c - \ln (B(u_k,v_k)^\top \mathbf{1}_n)$
\State $u_{k+1} \gets u_k$
\EndIf
\State $k \gets k +1$
\EndWhile
\State \Return $B(u_k,v_k)$.
\EndProcedure
\end{algorithmic}
\end{algorithm}

In Section \ref{sec:sink_def}, we provides some definitions used in Sinkhorn algorithm.
In Section \ref{sec:sink_bound_max_min}, we provides the bounds related to $u \in \R^n, v \in \R^n$.
In Section \ref{sec:sink_psi}, we define the potential function $\wt{\psi}$.
In Section \ref{sec:sink_upper_bound_psi}, we provide the upper bound of $\wt{\psi}$.
In Section \ref{sec:sink_iteration_bound}, we show the iteration complexity bound of the Sinkhorn's Algorithm.
In Section \ref{sec:sink_induction}, we provide the induction proof for the upper bound of the potential function.

\subsection{Definitions} \label{sec:sink_def}
We first introduce some definitions to simplify the derivations.
\begin{definition} \label{def:buv}
    We define matrix function $B: \R^{n} \times \R^n \rightarrow \R^{n \times n}$ as follows: for any given vectors $u, v \in \R^n$ 
    \begin{align*}
        B(u,v):=\diag(e^u) K \diag(e^v)
    \end{align*}
    where $\diag(a) \in \R^{n \times n}$ is the diagonal matrix with the vector $a \in \R^n$ on the diagonal and $K \in \R^{n \times n}$ is a matrix which is defined as 
\begin{align*}
K_{i,j} := \exp( -C_{i,j} /\gamma ).
\end{align*}
\end{definition}

\begin{definition}\label{def:psi}
    We define function $\psi : \R^{n} \times \R^n  \to \R$ as follows: for any given vectors $u,v \in \R^n$
    \begin{align*}
        \psi(u, v):=\mathbf{1}_n^\top B(u, v) \mathbf{1}_n-\langle u, r\rangle - \langle v, c\rangle, \
    \end{align*}
    where $B$ is defined in Definition \ref{def:buv}.
\end{definition}

We consider the Sinkhorn–Knopp algorithm (Algorithm \ref{alg:sinkhorn}), which solves 
the following minimization problem introduced in Lemma 2 of \cite{c13}:
\begin{align}\label{eq:min_pro}
    \min_{u, v \in \R^n}  \psi(u, v),
\end{align}
where $\psi$ is defined in Definition \ref{def:psi}.

Problem Eq.~\eqref{eq:min_pro} is the dual formulation to Eq.~\eqref{eq:reg_ot} as we choose $\mathcal{R}(X) = - H(X)$.

Here, we show the high level idea of proving the complexity of the Sinkhorn’s algorithm.

We first show how to get the bounds for $u_k,v_k$ and an optimal solution $(u_*,v_*)$ for Eq.~\eqref{eq:min_pro}. 

Next, we show that, for each iteration, $\psi(u_k ,v_k )$ is upper bounded by 
\begin{align*}
\|B(u_k ,v_k ) \mathbf{1}_n - r \|_1  + \|B(u_k ,v_k )^\top \mathbf{1}_n - c\|_1.
\end{align*}

Eventually, by using the bound of $\psi(u_k ,v_k )$, we show our result of the complexity result for the Sinkhorn's algorithm.

\begin{definition}\label{def:r}
    We define  $R$ as 
    \begin{align*}
        R:= & ~ - \ln ( K_{\min} \min_{i,j \in [n]} \{ r_i,c_j\} ),        .
    \end{align*}
    where 
    \begin{align*}
    K_{\min}:=&~ \min_{i,j \in [n]} K_{i,j} = e^{- \|C\|_{\infty/\gamma}}    
    \end{align*}
\end{definition}

\subsection{Bounded \texorpdfstring{$\max-\min$}{}} \label{sec:sink_bound_max_min}
We first present a tool related to the bounds for $u_k \in \R^n, v_k \in \R^n,u_* \in \R^n$ and $v_* \in \R^n$. 
\begin{lemma}\label{lem:bound_vku}
    Let $k \geq 0$ and $u_k \in \R^n, v_k \in \R^n$ be generated by Algorithm \ref{alg:sinkhorn} and $(u_*,v_*) \in \R^n \times \R^n$ be a solution of Eq.~\eqref{eq:min_pro}. Then 
    \begin{align}
        \max_{i \in [n]} u_{k,i} - \min_{i \in [n]} u_{k,i} \leq R,&~ \max_{j \in [n]} v_{k,j} - \min_{j \in [n]} v_{k,j} \leq R, \\
        \max_{i \in [n]} u_{*,i} - \min_{i \in [n]} u_{*,i} \leq R, &~ \max_{j \in [n]} v_{*,j} - \min_{j \in [n]} v_{*,j} \leq R, \notag
    \end{align}
    where $R$ is defined in Definition \ref{def:r}.

\end{lemma}

\begin{proof}
    First, we prove the bound for $u_k  \in \R^n$. As $u,v$ are initialized as $\mathbf{0}_n$, the inequality holds for $k=0$. Given $k-1$ is even, the variable $u$ is updated on the iteration $k-1$ and $B(u_k,v_k)\mathbf{1}_n=r$ by the algorithm construction. 
    
    Hence, for each $i\in [n]$  
    , we have
    \begin{align} \label{eq:e_k_1}
        e^{u_{k,i}} K_{\min} \langle \mathbf{1}_n, e^{v_k} \rangle 
        \leq & ~ \sum_{j=1}^n e^{e_{k,i}} K_{i,j} e^{v_{k,j}} \notag \\
        = & ~ [ B(u_k,v_k) (\mathbf{1}_n)_i] \notag \\ 
        = & ~ r_i \notag \\
        \leq & ~ 1 
    \end{align}
    where the first step follows from the definition of $K_{\min}$, the second step follows from the definition of $B$, the third step follows from $B(u_k,v_k)\mathbf{1}_n=r$ and the last step follows from the definition of probability simplex $r$.
    
    Hence, by reorganizing Eq.~\eqref{eq:e_k_1} we have
    \begin{align}\label{eq:max_uki_leq}
        \max_{i \in [n]} u_{k,i} \leq&~ -\ln(K_{\min} \langle \mathbf{1}_n,e^{v_k} \rangle).
    \end{align}
    On the other hand, since $0\leq K_{i,j} \leq 1$ 
    for each $i \in [n]$, 
    \begin{align*}
        &~ e^{u_{k,i}} \langle \mathbf{1}_n,e^{v_k} \rangle \\
        \geq&~ \sum_{j=1}^n e^{u_{k,i}} K_{i,j} e^{v_{k,j}} \\
        = & ~ [B(u_k,v_k) \mathbf{1}_n]_i \\
        =&~  r_i 
    \end{align*}
    where the first step follows from $K_{i,j} \leq 1$, the second step follows from the definition of $B$ and the last step follows from $B(u_k,v_k)\mathbf{1}_n=r$. 
    
    We also have
    \begin{align*}
         \min_{i \in [n]} u_{k,i} \geq&~\min_{i \in [n]} \ln(\frac{r_i}{\langle \mathbf{1}_n,e^{v_k} \rangle}) = \ln (\frac{\min_{i \in [n]} r_i}{\langle \mathbf{1}_n, e^{v_k} \rangle} ).
    \end{align*}
    The latter equality and Eq.~\eqref{eq:max_uki_leq} give
    \begin{align*}
        \max_{i \in [n]} u_{k,i} - \min_{i \in [n]} u_{k,i} \leq - \ln(K_{\min} \min_{i \in [n]} r_i) \leq R
    \end{align*}
\end{proof}

\subsection{Potential function \texorpdfstring{$\wt{\psi}$}{}}\label{sec:sink_psi}

To simplify derivations, we define $\wt{\psi}$ as follows:
\begin{definition}\label{def:wt_psi}
    We define $\wt{\psi}$ as
    \begin{align*}
    \wt{\psi}(u,v) := & ~ \psi(u,v) - \psi(u_*,v_*)
    \end{align*}
where the last step follows from the definition of $\psi$.
\end{definition}

\begin{claim}
We have
\begin{align*}
\wt{\psi}(u,v) = \langle \mathbf{1}_n, B(u,v) \mathbf{1}_n \rangle  - \langle \mathbf{1}_n, B(u_*,v_*) \mathbf{1}_n \rangle + \langle u_*-u,r \rangle + \langle v_*-v, c \rangle.
\end{align*}
\end{claim}
\begin{proof}
We can get
\begin{align*}
\wt{\psi}(u,v) 
= & ~ \psi(u,v) - \psi(u_*,v_*) \\
= & ~ \langle \mathbf{1}_n, B(u,v) \mathbf{1}_n \rangle  - \langle \mathbf{1}_n, B(u_*,v_*) \mathbf{1}_n \rangle + \langle u_*-u,r \rangle + \langle v_*-v, c \rangle.
\end{align*}
where the first step follows from the definition of $\wt{\psi}$, the second step follows from the definition of ${\psi}$.
\end{proof}

\subsection{Upper bounding for potential function} \label{sec:sink_upper_bound_psi}
Here, we provide a lemma which will be used later to bound the iteration complexity.
\begin{lemma}\label{lem:psi_uk_vk}

    Let $k\geq 1$ and $u_k,v_k \in \R^n$ be output of Algorithm \ref{alg:sinkhorn}. We denote $B_k := B(u_k,v_k)$. Then, we have
    \begin{align*}
        \wt{\psi} (u_k,v_k) \leq R \cdot (\|B_k \mathbf{1}_n - r\|_1 + \| B_k^\top \mathbf{1}_n - c\|_1).
    \end{align*}
\end{lemma}
\begin{proof}
    Given a fixed $k \geq 1$, for the following convex function of $(\hat{u},\hat{v})$
    \begin{align*}
        \langle \mathbf{1}_n,B(\hat{u},\hat{v}) \mathbf{1}_n \rangle - \langle \hat{u},B(u_k,v_k) \mathbf{1}_n \rangle - \langle \hat{v},B(u_k,v_k)^\top \mathbf{1}_n \rangle.
    \end{align*}
   The gradient of the convex function vanishes at $(\hat{u},\hat{v})=(u_k,v_k)$, so the point $(u_k,v_k)$ is its minimizer.
   
   Hence, 
    \begin{align}\label{eq:psi_upper_bound}
        \wt{\psi}(u_k,v_k) =&~ [\langle \mathbf{1}_n, B_k \mathbf{1}_n \rangle - \langle u_k, B_k \mathbf{1}_n \rangle - \langle v_k, B_k^\top \mathbf{1}_n \rangle] \notag \\
        &~ -[ \langle \mathbf{1}_n, B(u_*,v_*) \mathbf{1}_n \rangle - \langle u_*, B_k \mathbf{1}_n \rangle - \langle v_*, B_k^\top \mathbf{1}_n \rangle] \notag \\
        &~ +\langle u_k - u_*, B_k \mathbf{1}_n - r \rangle + \langle v_k - v_*, B_k^\top \mathbf{1}_n - c \rangle \notag \\
        \leq & ~ \langle u_k-u_*, B_k \mathbf{1}_n -r \rangle + \langle v_k - v_*, B_k^\top \mathbf{1}_n-c\rangle.
    \end{align}
    where the first step follows from the definition of $\wt{\psi}$. 
    Next, we bound the r.h.s of the inequality. For each iteration, we know that either $B_k \mathbf{1}_n = r$ or $B_k^\top \mathbf{1}_n = c$, so we have that $\langle \mathbf{1}_n, B_k \mathbf{1}_n \rangle = 1$ and $\langle \mathbf{1}_n, B_k \mathbf{1}_n - r\rangle = 0$. 
    
    Taking $a = 0.5 \cdot ( \max_{i \in [n]} u_{k,i} + \min_{i \in [n]} u_{k,i} )$.  
   Then, we have
    \begin{align*}
         \langle u_k, B_k \mathbf{1}_n - r \rangle 
        = & ~ \langle u_k - a\mathbf{1}_n, B_k \mathbf{1}_n -r \rangle \\
        \leq & ~ \|u_k - a\mathbf{1}_n\|_\infty \|B_k \mathbf{1}_n - r\|_1 \\
        =&~ 0.5 \cdot ( \max_{i \in [n]} u_{k,i} - \min_{i \in [n]} u_{k,i}) \|B_k \mathbf{1}_n - r\|_1 \\
        \leq & ~  \frac{R}{2} \|B_k \mathbf{1}_n - r\|_1.
    \end{align*}
    where the first step follows from $\langle \mathbf{1}_n, B_k \mathbf{1}_n - r\rangle = 0$ 
    , the second step follows from H\"{o}lder's inequality, the third step follows from the definition of $a$, and the last step follows from Lemma \ref{lem:bound_vku}. 
    
    Similarly, we bound $\langle -u_*, B_k \mathbf{1}_n - r \rangle, \langle v_k, B_k^\top \mathbf{1}_n - c \rangle$ and $\langle -v_*, B_k^\top \mathbf{1}_n - c \rangle$ in Eq.~\eqref{eq:psi_upper_bound} and complete the proof.
\end{proof}

\subsection{Iteration complexity bound}  \label{sec:sink_iteration_bound}
In this section, we show the iteration complexity bound for the Algorithm \ref{alg:sinkhorn}. 

\begin{theorem}\label{thm:upper_k}
   Given the cost matrix $C \in \R^{n \times n}$ and two simplex $r,c \in \R^n_+$, there is an algorithm (Algorithm \ref{alg:sinkhorn}) outputs $B(u_k,v_k)$ (Definition \ref{def:buv}) that satisfying 
    \begin{align*}
    \|B(u_k,v_k) \mathbf{1}_n - r\|_1 + \|B(u_k,v_k)^\top \mathbf{1}_n - c\|_1\leq \epsilon_0
    \end{align*}
    in the number of iterations $k$ satisfying 
    \begin{align*}
        k \leq 2+\frac{4R}{\epsilon_0}
    \end{align*}
\end{theorem}
\begin{proof}
    We first consider that $k \geq 1$ is even and define $B_k := B(u_k,v_k)$. We have
    \begin{align}\label{eq:temp_KL}
        &~ \psi(u_k, v_k) - \psi(u_{k+1}, v_{k+1}) \notag\\
        = & ~ \langle \mathbf{1}_n, B_k \mathbf{1}_n \rangle - \langle \mathbf{1}_n, B_{k+1} \mathbf{1}_n \rangle + \langle u_{k+1}-u_k,r \rangle + \langle v_{k+1}-v_k, c \rangle \notag\\
        =&~ \langle r,u_{k+1}-u_k \rangle\notag\\
        =&~ \langle r, \ln r - \ln (B_k \mathbf{1}_n)\rangle\notag\\
        =&~ \mathrm{KL}(r \| B_k \mathbf{1}_n )
    \end{align}

Then, we obtain
\begin{align} \label{eq:max_psi_uk}
    &~ \wt{\psi}(u_k, v_k)-\wt{\psi}(u_{k+1}, v_{k+1})\notag\\
    = & ~ \psi(u_k, v_k) - \psi(u_{k+1}, v_{k+1}) \notag\\
    =&~ \mathrm{KL} (r \|B_k \mathbf{1}_n) \notag\\
    \geq &~ \frac{1}{2} \|B_k \mathbf{1}_n - r\|_1^2 \notag\\
    \geq &~ \max \{ \frac{\wt{\psi}(u_k,v_k)^2}{2R^2},\frac{ \epsilon_0^2 }{2} \},
\end{align}
where the 1st step follows by the definition of $\wt{\psi}$, the 2nd step follows by Eq.~\eqref{eq:temp_KL}, the 3rd step follows by Pinsker's inequality and the last step follows by Lemma \ref{lem:psi_uk_vk} and $B_k^\top \mathbf{1}_n = c$.
For the last step, we also used that, as soon as the stopping criterion is not yet fulfilled and $B_k^\top \mathbf{1}_n = c$, $\|B_k \mathbf{1}_n - r\|_1^2 \geq \epsilon_0^2$. 
Similarly, when $k$ is odd, we can prove the same inequality. 

Given $\ell=\frac{2 R^2}{\wt{\psi} (u_1, v_1 )}$,  using Lemma~\ref{lem:induction}, we have for any $k \geq 1$
 \begin{align*}
\frac{\wt{\psi} (u_{k}, v_{k})}{2R^2} \leq \frac{1}{k+\ell-1}
\end{align*}

Thus,
\begin{align}\label{eq:k_geq_1}
    k \leq 1+\frac{2 R^2}{\wt{\psi} (u_k, v_k )}-\frac{2 R^2}{\wt{\psi} (u_1, v_1 )}
\end{align}
On the other hand,
\begin{align}\label{eq:psi_ukm_vkm}
    \wt{\psi} (u_{k+m}, v_{k+m} ) \leq \wt{\psi} (u_k, v_k )-\frac{ \epsilon_0^2 m}{2}, ~~ k, m \geq 0
\end{align}

Next, we use a switching strategy, parameterized by number $s \in (0, \wt{\psi}(u_1,v_1)]$, to combine Eq.~\eqref{eq:1_k_l} and Eq.~\eqref{eq:psi_ukm_vkm}.

First, by using Eq.~\eqref{eq:1_k_l}, we calculate the number of iterations needed to decrease $\wt{\psi}(u,v)$ from its initial value $\wt{\psi}(u_1,v_1)$ to a certain value $s$. Then, by applying Eq.~\eqref{eq:psi_ukm_vkm} and given $\wt{\psi}(u,v) \geq 0$ by its definition, we calculate the number of iterations required to further decrease $\wt{\psi}(u,v)$ from $s$ to zero. By minimizing the sum of these two estimates in $s \in (0,\wt{\psi}(u_1,v_1)]$, the total number of iterations $k$ satisfies the following
\begin{align*}
k \leq & ~ \min_{0<s \leq \wt{\psi} (u_1, v_1 )} (2+\frac{2 R^2}{s}-\frac{2 R^2}{\wt{\psi} (u_1, v_1 )}+\frac{2 s}{ \epsilon_0^2} ) \\
= & ~
\begin{cases}
2+\frac{4 R}{\epsilon_0}-\frac{2 R^2}{\wt{\psi} (u_1, v_1 )}, & \wt{\psi} (u_1, v_1 ) \geq R \epsilon_0, \\
2+\frac{2 \wt{\psi} (u_1, v_1 )}{ \epsilon_0^2}, & \wt{\psi} (u_1, v_1 )<R \epsilon_0 .
\end{cases}
\end{align*}
where the first step comes from Eq.~\eqref{eq:k_geq_1}, the first half of the last step comes from $a+b \geq 2\sqrt{ab}$ for $a\geq 0,~b\geq0$ and the second half follows from $s = \wt{\psi}(u_1,v_1)$.
In both cases, we have $k \leq 2+\frac{4R}{\epsilon_0}.$
\end{proof}

\subsection{Induction}\label{sec:sink_induction}

Here, we provide the induction proof for the upper bound of the potential function.
\begin{lemma}\label{lem:induction}
For all $k \geq 1$,
\begin{align*}
\frac{\wt{\psi} (u_{k}, v_{k})}{2R^2} \leq \frac{1}{k+\ell-1},
\end{align*}
where $\ell:=\frac{2 R^2}{\wt{\psi} (u_1, v_1 )}$ and $\wt{\psi}$ is defined in Definition \ref{def:psi}.
\end{lemma}
\begin{proof}
Our proof can be divided into two parts. At first, we consider the correctness of the in equalities above with $k = 1$.
Then, inducing over $k > 1$, the proof will be completed.

 {\bf Base Case.}
For $k=1$.
\begin{align*}
&~ \frac{\wt{\psi} (u_{1}, v_{1})}{2R^2}\\
= & ~ \frac{1}{\ell}\\
 = & ~ \frac{1}{k+\ell-1},
\end{align*}
where, the first step follows from the definition of $\ell$ and the last step follows from $k-1=0$. Hence, we have $\frac{\wt{\psi} (u_{k}, v_{k})}{2R^2} \leq \frac{1}{k+\ell-1}$ for k=1.

{\bf General case}
Suppose, 
\begin{align} \label{eq:general_case}
\frac{\wt{\psi} (u_{k}, v_{k})}{2R^2} \leq \frac{1}{k+\ell-1}
\end{align}
Then we can show
\begin{align}\label{eq:1_k_l}
    &~ \frac{\wt{\psi} (u_{k+1}, v_{k+1}  )}{2 R^2} \notag\\
    \leq&~ \frac{\wt{\psi} (u_k, v_k  )}{2 R^2}- (\frac{\wt{\psi} (u_k, v_k  )}{2 R^2}  )^2 \notag\\
    \leq & ~ \frac{1}{k + \ell - 1} - ( \frac{1}{k+\ell-1} )^2 \notag \\
    \leq&~ \frac{1}{k+\ell},
\end{align}
where the first step follows from Eq.~\eqref{eq:max_psi_uk}, and the second step follows from Eq.~\eqref{eq:general_case} and the property of function $f(x ) = x - x^2$ (which is $f(y) \leq f(z)$ if $y \leq z \leq 1/2$), the last step follows from $\frac{1}{A} - \frac{1}{A^2} \leq \frac{1}{A+1}$ for any integer $A \geq 2$.
By induction, the proof is completed.
\end{proof}

\section{Running Time with small treewidth setting}\label{sec:ot_treewidth}
In Section \ref{sec:treewidth_implicit_k}, we introduce the implicit form $K$.
In Section \ref{sec:treewidth_sinkhorn_time}, we provided the faster Sinkhorn' Algorithm with small treewidth.
In Section \ref{sec:treewidth_round_correctness}, we show the correctness of our rounding algorithm.
In Section \ref{sec:treewidth_round_time}, we show the running time needed for our rounding algorithm.
In Section \ref{sec:treewidth_otdistance_time}, we provide the running time for approximating the OT distance by using the faster Sinkhorn's Algorithm.

\subsection{Implicit form of \texorpdfstring{$K$}{}}\label{sec:treewidth_implicit_k}

Here we introduce the implicit form of $K$ to make use of the small treewidth setting.

\begin{lemma}\label{lem:spa_l_a}
    We assume $C=MM^\top \in \R^{n \times n}$, where $M \in \R^{n \times d}$ has treewidth $\tau$. Given $A := K - D$, where $D_{i,j}:= 1$ for $i,j \in [n]$ and $K$ is defined in Definition \ref{def:buv}, the Cholesky factor $L_A$ for $A = L_A L_A^\top$ is $\tau$-sparse in columns. 
\end{lemma}
\begin{proof}
    Given $C=MM^\top$ and $M$ has treewidth $\tau$, the Cholesky factor $L_C$ for $C = MM^\top =L_ML_M^\top$ is $\tau$ sparse in column by using Lemma \ref{lem:fast_cholesky}. As 
    \begin{align*}
        A_{i,j} = e^{-C_{i,j} /\gamma} - 1,
    \end{align*}
    we have $A_{i,j} = 0$ when $C_{i,j} = 0$. Hence, matrix $A$ is as sparse as matrix $C$. We have that the Cholesky factor $L_A$ for $A = L_A L_A^\top$ is as sparse as $L_M$. As $L_M$ is $\tau$-sparse, we complete the proof.
\end{proof}
 
\begin{algorithm}[!ht]
\caption{Sinkhorn’s Algorithm with small treewidth
}\label{alg:sinkhorn_treewidth}
\begin{algorithmic}[1]
\Procedure{SinkhornAlgorithm}{$r \in \R^n,c \in \R^n,\epsilon_0 \in (0,1)$}  \Comment{Theorem \ref{thm:per_sinkhorn_treewidth}}
\State \Comment{Accuracy $\epsilon_0$}
\State $k \gets 0 $,
\State $u_0 \gets 0$
\State $v_0 \gets 0$
\State $w \gets \mathbf{1}_n$
\State $x_0 \gets  e^{-C_{i,j} /\gamma}  \mathbf{1}_n$
\State $y_0 \gets  (e^{-C_{i,j} /\gamma}) ^\top \mathbf{1}_n$
\State Implicitly form $D = w w^\top$
\State Implicitly form $A \in \R^{n \times n}$, where $A_{i,j} = e^{-C_{i,j} /\gamma} - 1$ 
\State \Comment{Explicitly writing down $A$ requires $n^2$, however, we never need to explicitly write down $A$. Knowing the exact formulation of $A$ is enough to do the Cholesky decomposition} 
\State $L \gets$ Cholesky decomposition matrix for $A$ i.e., $A = L_A L_A^\top $  \Comment{$O(n \tau^2)$, Lemma \ref{lem:cholesky_time} }
\While{$\|x_k -r\|_1 + \|  y_k -c\|_1 \ge \epsilon_0$}
\If {$k \mod 2 =0$}
\State $u_{k+1} \gets u_k + \ln r - \ln (x_k)$
\State $v_{k+1} \gets v_k$
\Else
\State $v_{k+1} \gets v_k + \ln c - \ln (y_k)$
\State $u_{k+1} \gets u_k$
\EndIf
\State $x_k \gets (\diag(e^{u_k}) (L_A L_A^\top) \diag(e^{v_k})  +\diag(e^{u_k}) D\diag(e^{v_k}) ) {\bf 1}_n$
\State $y_k \gets (\diag(e^{u_k}) (L_A L_A^\top) \diag(e^{v_k})  +\diag(e^{u_k}) D\diag(e^{v_k}) )^\top {\bf 1}_n$
\State $k \gets k +1$
\EndWhile
\State \Return $u_k, v_k, L_A, w$ \Comment{We return $B(u_k,v_k)$ in a implicit way, i.e., $B(u_k,v_k) = \diag(e^{u_k}) (L_A L_A^\top) \diag(e^{v_k})  +\diag(e^{u_k}) (w w^\top) \diag(e^{v_k})$}. 
\EndProcedure
\end{algorithmic}
\end{algorithm}

\subsection{Running time of Sinkhorn with small treewidth}\label{sec:treewidth_sinkhorn_time}

This section is to prove the running time of Algorithm \ref{alg:sinkhorn_treewidth}. 
\begin{theorem}[Running time of Algorithm \ref{alg:sinkhorn_treewidth}] \label{thm:per_sinkhorn_treewidth}
    Given the cost matrix $C \in \R^{n \times n}$ with small treewidth $\tau$ and two simplex $r,c \in \R^{n}_+$, there is an algorithm (Algorithm \ref{alg:sinkhorn_treewidth}) takes $O(n \tau)$ for each iteration and $O( n \tau^2)$ for initialization to output 
    \begin{itemize}
    \item a lower triangular matrix $L_A$
    \item vectors $u,v, w \in \R^n$
    \end{itemize}
    such that $B(u_k,v_k) \in \R^{n \times n}$  can be constructed (implicitly) by 
    \begin{align*}
    B(u_k,v_k) = \diag(e^{u_k}) (L_A L_A^\top) \diag(e^{v_k})  +\diag(e^{u_k}) (w w^\top) \diag(e^{v_k})
    \end{align*}
    satisfying 
    \begin{align*}
    \|B(u_k,v_k) \mathbf{1}_n - r\|_1 + \|B(u_k,v_k)^\top \mathbf{1}_n - c\|_1\leq \epsilon_0.
    \end{align*}
    \end{theorem}
    \begin{proof}
    The running time for each step is shown as follows:
    \begin{itemize}
        \item Writing down cost matrix $C \in \R^{n \times n}$ takes $O(n\tau)$ time as $\nnz(C) = n \tau$ by using Claim \ref{cla:nnz_c}.
        \item Implicitly write down matrix $D \in \R^{n \times n}$, this takes $O(n)$ time since $D \in \R^{n \times n}$ is a rank-$1$ matrix. 
        \item Initializing $x_0$ and $y_0$ takes $O(n \tau)$ as $\nnz(C) = n \tau$.
        \item Using Lemma \ref{lem:spa_l_a}, we know $L_A$ is $\tau$-sparse in column. Then, calculating the Cholesky decomposition for $A$ takes $O(n \tau^2)$ time using Lemma \ref{lem:cholesky_time}.
        \item Calculating  $\diag(e^{u_k}) (L_A L_A^\top) \diag(e^{v_k})$ takes $O(n \tau)$ time as $L_A$ is $\tau$-sparse in column.
        \item Calculating  $\diag(e^{u_k}) D\diag(e^{v_k})$ takes $O(n)$ time as matrix $D$ is a rank-1 matrix.
        \item Updating $u \in \R^n, v \in \R^n$ takes $O(n)$ time.
    \end{itemize}
    Hence, the initialization time for Algorithm \ref{alg:sinkhorn_treewidth} is $O(n \tau^2)$ and the per iteration running time is $O(n \tau)$.
\end{proof}

\begin{algorithm}[!ht]
\caption{Approximate OT by Sinkhorn
}\label{alg:ot_dis}
\begin{algorithmic}[1]
\Procedure{ApproxOT}{$\epsilon$} \Comment{Theorem \ref{thm:runningtime_ot}}  
\State \Comment{Accuracy $\epsilon$}
\State $\gamma \gets \frac{\epsilon}{4 \ln n}$
\State $\epsilon_0 \gets \frac{\epsilon}{8 \|C\|_\infty}$
\State \Comment{Find $\wt{r}, \wt{c} \in \Delta^n$ s.t. $\|\wt{r}-r\|_1 \leq \epsilon_0 / 4, \|\wt{c}-c\|_1 \leq \epsilon_0 / 4$ and $\min_{i\in [n]} \wt{r}_i \geq \epsilon_0 / (8n), \min_{j\in [n]} \wt{c}_j \geq \epsilon_0 / (8n)$.} 
\State $(\wt{r}, \wt{c}) \gets (1-\frac{\epsilon_0}{8} ) ((r, c)+\frac{\epsilon_0}{n (8-\epsilon_0 )}(\mathbf{1}_n, \mathbf{1}_n) )$\label{line:find_r_c}

\State $(u,v,L,w) \gets \textsc{SinkhornAlgorithm}(\wt{r},\wt{c}$,$\epsilon_0 / 2$) \Comment{Algorithm \ref{alg:sinkhorn_treewidth}} \label{line:calculate_b}
\State \Comment{Note that $u,v,L,w$ is an implicit representation of $B$, i.e., $\diag(e^{u_k}) (L_A L_A^\top) \diag(e^{v_k})  +\diag(e^{u_k}) (w w^\top) \diag(e^{v_k})$}
\State  $(p$, $q$, $X, Y, w, u, v)$ $\gets \textsc{Round}(u,v,L,w,r,c)$ \Comment{Algorithm \ref{alg:round}} 
\label{line:find_x}
\State \Return $(p$, $q$, $X, Y, L, w, u, v)$   \Comment{We return $\hat{X}$ in an implicit way, i.e., $\hat{X} := XBY + p q^\top / \|p\|_1$ }
\EndProcedure
\end{algorithmic}
\end{algorithm}
\begin{algorithm}[!ht]
\caption{Rounding of the projection of $B$ on $\mathcal{U}$}\label{alg:round}
\begin{algorithmic}[1]
\Procedure{Round}{$u \in \R^n, v \in \R^n,  L \in \R^{n \times n} ,w \in \R^{n},r \in \R^n, c \in \R^n$} \Comment{Lemma \ref{lem:round}   
}
\State \Comment{$L$ is a lower triangular matrix that only has $O(n \tau)$ nonzeros}
\State \Comment{We never explicit write $B$. $B$ can implicitly represented by $\diag(e^{u}) (L_A L_A^\top) \diag(e^{v})  +\diag(e^{u}) (w w^\top) \diag(e^{v})$}
\State $X \gets \diag(x)$ with $x_i = \min\{\frac{r_i}{r_i(B)}, 1\}$  \Comment{$r(B) := B \mathbf{1}_n$, $X \in \R^{n \times n}$}
\State $B_0 \gets XB$ \Comment{We only implicitly construct $B_0$}
\State $Y \gets \diag(y)$ with $y_j =  \min\{\frac{c_j}{c_j(B_0)} , 1\}$ \Comment{$c(B_0) := B_0^\top \mathbf{1}_n$}
\State $B_1 \gets B_0 Y$ \Comment{We only implicitly construct $B_1$}
\State $p \gets r-B_1 \mathbf{1}_n, q \gets c -B_1^\top \mathbf{1}_n$ 
\State \Return $p$, $q$, $X, Y, w, u, v$ \Comment{We return $G$ in an implicit way, i.e., $G := XBY + p q^\top / \|p\|_1$ }
\EndProcedure
\end{algorithmic}
\end{algorithm}

\subsection{ Correctness of rounding algorithm}\label{sec:treewidth_round_correctness}
We first show the correctness of our rounding algorithm (Algorithm \ref{alg:round}).
\begin{lemma}[An improved version of of Lemma 7 in \cite{anr17}]\label{lem:round_correct}
Given $r,c \in \triangle_n$, $B \in \R_+^{n \times n}$, $u,v \in \R^{n}$ and $r,c \in \R^n$, there is an algorithm (Algorithm \ref{alg:round}) outputs 
\begin{itemize}
\item a diagonal matrix $X \in \R^{n \times n}$
\item a diagonal matrix $Y \in \R^{n \times n}$
\item a lower triangular matrix $L_A$
\item vectors $u,v, w \in \R^n$
\item vectors $p \in \R^n$, $q \in \R^n$
\end{itemize}
such that $G \in \mathcal{U}_{r,c}$  can be constructed (implicitly) by 
\begin{align*}
\hat{X} = X ( \diag(e^u) L_A L_A^\top \diag(e^v) + \diag(e^u) (w w^\top) \diag(e^v) ) Y  + p q^\top / \|p\|_1
\end{align*}
that satisfying 
\begin{align*}
     \|G-B\|_1 \leq 2(\|B \mathbf{1}_n -r\|_1+\|B^\top \mathbf{1}_n-c\|_1).
\end{align*}
\end{lemma}
\begin{proof}
Let $G$ be the output of Algorithm \ref{alg:round}. As matrix $B_1$ are nonnegative, and the output $q$ and $p$ are both negative, with $\| p\|_1 = \|q \|_1 = 1 - \|B_1\|_1$, matrix $G$ are nonnegative and  
\begin{align}\label{eq:r_g_r_b1}
    r(G) =&~ r(B_1)+r (p q^{\top} / \|p \|_1 )\notag\\
     =&~r(B_1)+p \notag\\
     =&~ r,
\end{align}
where we denote $r(A) := A \mathbf{1}_n, c(A) := A^\top \mathbf{1}_n$ and the first two step comes from the definition of $r$ and the last step comes from $p = r - B_1 \mathbf{1}_n$.
Similarly, we have $c(G) = c $. Therefore, we have $G \in \mathcal{U}_{r, c}$.

Next, we denote $\Delta:=$ $\|B\|_1- \| B_1 \|_1$ and prove the $\ell_1$ bound between the matrix $B$ and matrix $G$. We first remove mass from a row of $B$ when $r_i(B) \geq r_i$, and then, we remove mass from a column when $c_j (B_0 ) \geq c_j$. Now, we have
\begin{align}\label{eq:delta_sum}
\Delta=\sum_{i=1}^n (r_i(B)-r_i )_{+}+\sum_{j=1}^n (c_j (B_0 )-c_j )_{+} .
\end{align}
Then, we show the analysis of Eq.~\eqref{eq:delta_sum}. First, for the left sum of Eq.~\eqref{eq:delta_sum}, we have
\begin{align*}
\sum_{i=1}^n (r_i(B)-r_i )_{+}=\frac{1}{2} (\|r(B)-r\|_1+\|B\|-1 ) .
\end{align*}
For the second sum in Eq.~\eqref{eq:delta_sum}.
\begin{align*}
    \sum_{j=1}^n (c_j (B_0 )-c_j )_{+} \leq \sum_{j=1}^n (c_j(B)-c_j )_{+} \leq\|c(B)-c\|_1
\end{align*}
 where the first step comes from the fact that the vector $c(B)$ is entrywise larger than $c (B_0 )$ and the last step comes from the definition of $c$.

Therefore we conclude

\begin{align}
\|G-B\|_1 \leq & ~ \Delta+ \|p q^{\top} \|_1 / \|p \|_1 \notag\\
= & ~\Delta+1- \| B_1 \|_1 \notag\\
= & ~2 \Delta+1-\|B\|_1 \notag\\
\leq & ~\|r(B)-r\|_1+2\|c(B)-c\|_1 \\
\leq & ~ 2 (\|r(B)-r\|_1+\|c(B)-c\|_1 ) \notag
\end{align}
where the first step comes from the definition of $\Delta$, the second step comes from the fact that $\| p\|_1 = \|q \|_1 = 1 - \|B_1\|_1$, the third step comes from the definition of $\Delta$, the fourth step comes from Eq.~\eqref{eq:delta_sum} and the last step comes from reorganization.
Now we complete the proof.
\end{proof}

\subsection{ Running time of rounding algorithm}\label{sec:treewidth_round_time}
Next, we show the running time needed for the rounding algorithm (Algorithm \ref{alg:round}).

\begin{lemma}[An improved version of of Lemma 7 in \cite{anr17}]\label{lem:round}
Given $r,c \in \triangle_n$, $B \in \R_+^{n \times n}$, $u,v \in \R^{n}$ and $r,c \in \R^n$, there is an algorithm (Algorithm \ref{alg:round}) outputs 
\begin{itemize}
\item a diagonal matrix $X \in \R^{n \times n}$
\item a diagonal matrix $Y \in \R^{n \times n}$
\item a lower triangular matrix $L_A$
\item vectors $u,v, w \in \R^n$
\item vectors $p \in \R^n$, $q \in \R^n$
\end{itemize}
such that $G \in \mathcal{U}_{r,c}$  can be constructed (implicitly) by 
\begin{align*}
\hat{X} = X ( \diag(e^u) L_A L_A^\top \diag(e^v) + \diag(e^u) (w w^\top) \diag(e^v) ) Y  + p q^\top / \|p\|_1
\end{align*}
that satisfying 
\begin{align*}
     \|G-B\|_1 \leq 2(\|B \mathbf{1}_n -r\|_1+\|B^\top \mathbf{1}_n-c\|_1)
\end{align*}
in $O(n \tau)$ time.
\end{lemma}
\begin{proof}

The running time for each step is shown as follows: %
\begin{itemize}
    \item Calculating $r(B)$ takes $O(n \tau)$ time. Given 
    \begin{align*}
        r(B) = B \mathbf{1}_n = \diag(e^{u_k}) (L_A L_A^\top) \mathbf{1}_n \diag(e^{v_k})  +\diag(e^{u_k}) (w w^\top) \mathbf{1}_n \diag(e^{v_k}),
    \end{align*}
    calculating $L_A (L_A^\top \mathbf{1}_n)$ takes $ O(n \tau)$, as $\nnz(L_A) = n \tau$. As $w = \mathbf{1}_n$, calculating $(w w^\top) \mathbf{1}_n$ takes $O(n)$.
    \item Calculating $X = \diag(x)$ with $x_i = \min\{\frac{r_i}{r_i(B)}, 1\}$ takes $O(n)$ time. 
    \item For $B_0 = XB$, we remark that $B_0$ is not explicitly written down. It is implicitly represented by $L_A,w,u,v,X$.
    \item Similarly, we can calculate $Y$ in $O(n)$ time and implicitly write down $B_1$.
    \item We have
     \begin{align*}
        B_1 \mathbf{1}_n = X B Y =  \diag(e^{u_k})X (L_A L_A^\top)Y \mathbf{1}_n \diag(e^{v_k})  + \diag(e^{u_k}) X(w w^\top)Y  \mathbf{1}_n \diag(e^{v_k}).
    \end{align*}
    For any diagonal matrix $M$, $M \cdot L_A$ is as sparse as $L_A$ and it takes $O(n \tau)$ to compute it. Therefore, computing $P = \diag(e^{u_k})X (L_A L_A^\top)Y  \diag(e^{v_k})$ takes $O(n \tau)$ time and $P$ is $ n \tau$-sparse. Then, we compute $P \mathbf{1}_n$, which takes $O( n \tau)$ time. Hence, updating $p$ takes $O( n \tau)$ time.
    \item Similarly, updating $q$ takes $O( n \tau)$ time.
    \item For matrix $G$, it is returned in an implicit way. We use $p$, $q$, $X, Y, w, u, v$ to represent it.
\end{itemize}
Therefore, the total running time is $O(n \tau)$.
\end{proof}

\subsection{Running time of OT Distance by Sinkhorn }\label{sec:treewidth_otdistance_time}
The core of our OT algorithm is the entropic penalty
\begin{align}\label{eq:penalty}
    X_\gamma := \arg\min_{X \in \mathcal{U}_{r,c}}\langle X,C \rangle + \gamma \cdot \mathcal{R}(X) .
\end{align}
The solution to Eq.~\eqref{eq:penalty} can be characterized explicitly by analyzing its first-order conditions for optimality.

Now we apply the result of the previous subsection to derive a complexity estimate for finding $\hat{X} \in \mathcal{U}(r,c)$ satisfying Eq.~\eqref{eq:app_ot}. The procedure for approximating the OT distance by the Sinkhorn’s algorithm is listed as Algorithm \ref{alg:ot_dis}.

\begin{theorem}\label{thm:runningtime_ot}

There is an algorithm (Algorithm \ref{alg:ot_dis}) takes cost matrix $C \in \R^{n \times n}$, two $n$-dimensional simplex $r,c$ as inputs and outputs 
\begin{itemize}
\item a diagonal matrix $X \in \R^{n \times n}$
\item a diagonal matrix $Y \in \R^{n \times n}$
\item a lower triangular matrix $L_A$
\item vectors $u,v, w \in \R^n$
\item vectors $p \in \R^n$, $q \in \R^n$
\end{itemize}
such that $\hat{X} \in \mathcal{U}(r,c)$  can be constructed (implicitly) by 
\begin{align*}
\hat{X} = X ( \diag(e^u) L_A L_A^\top \diag(e^v) + \diag(e^u) (w w^\top) \diag(e^v) ) Y  + p q^\top / \|p\|_1
\end{align*}
that satisfying Eq.~\eqref{eq:app_ot} in 
    \begin{align*}
        O( n \tau^2 + \epsilon^{-2} n\tau \|C\|_\infty^2 \ln n  )
    \end{align*}
    time.
\end{theorem}
\begin{remark}
If we don't care about the output format to be lower-triangular matrix, then the additive term $n\tau^2$ can be removed.
\end{remark}

\begin{proof}
    Let $X_* \in \arg\min_{X \in \mathcal{U}_{r,c}} \langle P,C \rangle$ be an optimal solution to the original OT program.
    
    We first show that $\langle B , C\rangle$ is not much larger than $\langle X_*,C \rangle$.
    
    Since $B = MAN \in \R^{n \times n}$ 
    for positive diagonal matrices $M,N \in \R^{n \times n}_+$,  
    Lemma \ref{lem:unique_min} implies $B$ is the optimal solution to 
    \begin{align}\label{eq:optimal_solutlion_to}
        \arg\min_{X \in \mathcal{U}_{r,c}}\langle X,C \rangle + \gamma \mathcal{R}(X) .
    \end{align}
    By Lemma \ref{lem:round}, there exists a matrix $X_0 \in \mathcal{U}_{B \mathbf{1}_n,B^\top \mathbf{1}_n}$ (Definition~\ref{def:u_r_c}) 
    such that 
    \begin{align}\label{eq:x0_xstar}
         \|X_0-X_* \|_1 \leq 2 ( \|B \mathbf{1}_n-r \|_1+ \|B^\top \mathbf{1}_n-c \|_1 ).
    \end{align}
    Moreover, since $B \in \R^{n \times n}$ is an optimal solution of Eq.~\eqref{eq:optimal_solutlion_to}, we have
    \begin{align}\label{eq:b_c_gamma}
        \langle B, C\rangle+\gamma \mathcal{R}(B)\leq \langle X _0, C \rangle+\gamma \mathcal{R}(X_0).
    \end{align}
    Thus, we have 
    \begin{align}\label{eq:B_C_X_C}
    \langle B, C\rangle-  \langle X_*, C  \rangle = & ~\langle B, C\rangle-  \langle X_0, C  \rangle+  \langle X_0, C  \rangle-  \langle X_*, C  \rangle \notag\\
    = & ~\langle B, C\rangle-  \langle X_0, C  \rangle+ \|X_0-X_* \|_1  \|C\|_{\infty} \notag\\
    \leq & ~ \gamma  (H(B)-H  (X_0  )  )+ \|X_0-X_* \|_1  \|C\|_{\infty}  \notag\\
    \leq & ~ \gamma  (H(B)-H  (X_0  )  )+2  (  \|B \mathbf{1}_n-r  \|_1+  \|B^\top \mathbf{1}_n-c  \|_1  )\|C\|_{\infty} \notag\\
    \leq & ~ 2 \gamma \ln n+2  (  \|B \mathbf{1}_n-r  \|_1+  \|B^\top \mathbf{1}_n-c  \|_1  )\|C\|_{\infty}
    \end{align}
where the first step follows from reorganization, the second step follows from H\"{o}lder's inequality (Lemma \ref{lem:holder}), the third step follows from Eq.~\eqref{eq:b_c_gamma} and $\mathcal{R}(X) = - H(X)$, the fourth step follows from Eq.~\eqref{eq:x0_xstar} and the last step follows from the fact that $0 < H(B),H(X_0) \leq 2 \ln n$. 

Lemma \ref{lem:round} implies that the output $\hat{X}$ of Algorithm \ref{alg:round} satisfies the inequality 
\begin{align}\label{eq:b_hat_x_1}
\|B-\hat{X}\|_1 \leq 2 ( \|B \mathbf{1}_n-r \|_1+ \|B^\top \mathbf{1}_n-c \|_1 ).
\end{align}

Recall that $\hat{X}$ is the output of Algorithm \ref{alg:ot_dis}, $X_*$ is a solution to the OT problem Eq.~\eqref{eq:app_ot} and $B$ is the matrix obtained in line \ref{line:calculate_b} of Algorithm \ref{alg:ot_dis}. We have
\begin{align}\label{eq:upper_c_x}
    \langle\hat{X}, C\rangle = &~ \langle \hat{X}-B, C \rangle + \langle B,C \rangle \notag\\
    \leq &~ \|\hat{X}-B\|_1 \|C\|_\infty + \langle B,C \rangle \notag \\
    \leq &~  2 ( \|B \mathbf{1}_n-r \|_1+ \|B^\top \mathbf{1}_n-c \|_1 ) \|C\|_\infty + \langle B,C \rangle \notag \\
    \leq &~ \langle  X_*, C\rangle+2 \gamma \ln n+4 ( \|B \mathbf{1}_n-r \|_1+ \|B^\top \mathbf{1}_n-c \|_1 )\|C\|_{\infty} .
\end{align}
 where the first step follows from reorganization, the second step follows from H\"{o}lder's inequality, the third step follows from Eq.~\eqref{eq:b_hat_x_1} and the last step follows from Eq.~\eqref{eq:B_C_X_C}.

    At the same time, we have 
    \begin{align*}
        & ~ \|B \mathbf{1}_n-r\|_1+ \|B^\top \mathbf{1}_n-c \|_1 \\
        \leq & ~\|B \mathbf{1}_n-\wt{r}\|_1+\|\wt{r}-r\|_1+ \|B^\top \mathbf{1}_n-\wt{c} \|_1+\|\wt{c}-c\|_1 \\
        \leq & ~  \epsilon_0
    \end{align*}
    where the first step follows from the definition of $\ell_1$-norm and the last step follows from $\|B \mathbf{1}_n -r\|_1 + \|B^\top \mathbf{1}_n-c\|_1 \le \epsilon_0$ (output of Algorithm \ref{alg:sinkhorn_treewidth}) and the definitions of $\wt{r}$ and $\wt{c}$.

    Setting $\gamma=\frac{\epsilon }{ 4 \ln n }$ and $\epsilon_0 = \frac{\epsilon}{8 \|C\|_\infty}$, we obtain from the above inequality and Eq.~\eqref{eq:upper_c_x} that $\hat{X}$ satisfies inequality Eq.~\eqref{eq:app_ot}.
    
    Next, we show complexity of Algorithm \ref{alg:ot_dis}. When $\epsilon_0$ is sufficiently small, the number of iterations of the Sinkhorn’s algorithm in line \ref{line:calculate_b} of Algorithm \ref{alg:ot_dis} is $O({R}/{\epsilon_0})$, by using Theorem \ref{thm:upper_k}. According to Definition \ref{def:r}, we have
    \begin{align*}
    R= & ~ -\ln  (K_{\min} \min_{i, j \in [n]} \{\wt{r}_i, \wt{c}_j \} ) \\
    = & -\ln  (e^{-\|C\|_{\infty} / \gamma} \min_{i, j \in [n]} \{\wt{r}_i, \wt{c}_j \} ) \\
    \leq & ~ \frac{\|C\|_{\infty}}{\gamma}-\ln  (\frac{\epsilon_0}{8 n} ) ,
    \end{align*}
    where the first step follows from the definition of $R$,the second step follows from the definition of $K_{\min}$, the last step follows from the condition of $\wt{r}_i, \wt{c}_j$ in line \ref{line:find_r_c} of Algorithm \ref{alg:ot_dis}.

Since $\gamma=\frac{\epsilon}{4 \ln n}$ and $\epsilon_0=\frac{\epsilon}{8\|C\|_{\infty}}$, we have that 
\begin{align*}
R=O (\epsilon^{-1} \|C\|_{\infty} \ln n ).
\end{align*}

As the number of iteration for Algorithm \ref{alg:ot_dis} is $O({R}/{\epsilon_0})$, we conclude that the total number of Sinkhorn's algorithm iterations is bounded by $O (\epsilon^{-2} \|C\|_{\infty}^2 \ln n )$.

Obviously, $\wt{r} \in \R_+^n$ and $\wt{c} \in \R^n_+$ in line \ref{line:find_r_c} of Algorithm \ref{alg:ot_dis} can be found in $O(n)$ time.

Since each iteration of the Sinkhorn's algorithm requires $O (n \tau)$ time and the initialization takes $O(n\tau^2)$ time as shown in Theorem \ref{thm:per_sinkhorn_treewidth}, the total complexity of Algorithm \ref{alg:ot_dis} is 
\begin{align*}
    O (n \tau^2 +  \epsilon^{-2} n \tau \|C\|_{\infty}^2 \ln n ).
\end{align*}
\end{proof}

\section{Symmetric}
\label{sec:sym}

In this section, we provide an algorithm (Algorithm \ref{alg:ot_dis_sym}) to solve the OT problem in $ O( \epsilon^{-2} n\tau \|C\|_\infty^2 \ln n  )$ time, given the two distribution are identical, i.e., $c=r$.
\begin{definition}
    Given the symmetric OT problem $\arg\min_{X \in \mathcal{U}_{r }}\langle X  \rangle $, we define
    \begin{align*}
         \mathcal{U}_{r } = \{X \in \R^{n \times n}_+: X \mathbf{1}_n=r, X^\top \mathbf{1}_n = r\}
    \end{align*}
    where $\mathbf{1}_n$ is the all-ones vector in $\R^n$ , $C\in \R^{n\times n}_+$ is a given cost matrix, and $r \in \R ^n $ are given vectors with positive entries that sum to one.
\end{definition}

\begin{algorithm}[!ht]
\caption{Sinkhorn’s Algorithm for symmetric distribution with small treewidth}\label{alg:sinkhorn_treewidth_sym}
\begin{algorithmic}[1]
\Procedure{SinkhornAlgorithmSym}{$r,\epsilon_0 \in (0,1)$}  \Comment{Theorem \ref{thm:per_sinkhorn_treewidth_sym}}
\State \Comment{Accuracy $\epsilon_0$}
\State $k \gets 0 $,
\State $u_0 \gets 0$
\State $v_0 \gets 0$
\State $w \gets \mathbf{1}_n$
\State $x_0 \gets  e^{-C_{i,j} /\gamma}  \mathbf{1}_n$
\State $y_0 \gets  (e^{-C_{i,j} /\gamma}) ^\top \mathbf{1}_n$
\State Implicitly form $D = w w^\top$
\State Implicitly form $A \in \R^{n \times n}$, where $A_{i,j} = e^{-C_{i,j} /\gamma} - 1$ 
\State \Comment{Explicitly writing down $A$ requires $n^2$, however, we never need to explicitly write down $A$. Knowing the exact formulation of $A$ is enough to do the Cholesky decomposition} 
\State $L \gets$ Cholesky decomposition matrix for $A$ i.e., $A = L_A L_A^\top $  \Comment{$O(n \tau^2)$, Lemma \ref{lem:cholesky_time} }
\While{$\|x_k -r\|_1 \ge \epsilon_0$}
\State $u_{k+1} \gets u_k + \ln r - \ln (x_k)$
\State $x_k \gets (\diag(e^{u_k}) (L_A L_A^\top) \diag(e^{u_k})  +\diag(e^{u_k}) D\diag(e^{u_k}) ) {\bf 1}_n$ 
\State $k \gets k +1$
\EndWhile
\State \Return $u_k, L_A, w$ \Comment{We return $B(u_k)$ in a implicit way, i.e., $B(u_k) = \diag(e^{u_k}) (L_A L_A^\top) \diag(e^{u_k})  +\diag(e^{u_k}) (w w^\top) \diag(e^{u_k})$}. 
\EndProcedure
\end{algorithmic}
\end{algorithm}

We first provide the running time of the Sinkhorn's algorithm (Algorithm \ref{alg:sinkhorn_treewidth_sym}) for symmetric case.
\begin{theorem}[Running time of Algorithm \ref{alg:sinkhorn_treewidth_sym}] \label{thm:per_sinkhorn_treewidth_sym}
    Given the cost matrix $C \in \R^{n \times n}$ with small treewidth $\tau$ and a simplex $r \in \R^{n}_+$, there is an algorithm (Algorithm \ref{alg:sinkhorn_treewidth_sym}) takes $O(n \tau)$ for each iteration and $O( n \tau^2)$ for initialization to output 
    \begin{itemize}
    \item a lower triangular matrix $L_A$
    \item vectors $u, w \in \R^n$
    \end{itemize}
    such that $B(u_k) \in \R^{n \times n}$  can be constructed (implicitly) by 
    \begin{align*}
    B(u_k) = \diag(e^{u_k}) (L_A L_A^\top) \diag(e^{u_k})  +\diag(e^{u_k}) (w w^\top) \diag(e^{u_k})
    \end{align*}
    satisfying 
    \begin{align*}
    \|B(u_k) \mathbf{1}_n - r\|_1 \leq \epsilon_0.
    \end{align*}
    \end{theorem}
\begin{proof}
    Similar to the proof of Theorem \ref{thm:per_sinkhorn_treewidth}, here the two distribution are identical, i.e., $c=r$.
\end{proof}

\begin{algorithm}[!ht]
\caption{Approximate OT by Sinkhorn for symmetric distribution}\label{alg:ot_dis_sym}
\begin{algorithmic}[1]
\Procedure{ApproxOTSym}{$\epsilon$} \Comment{Theorem \ref{thm:runningtime_ot_sym}}  
\State \Comment{Accuracy $\epsilon$}
\State $\gamma \gets \frac{\epsilon}{4 \ln n}$
\State $\epsilon_0 \gets \frac{\epsilon}{8 \|C\|_\infty}$
\State \Comment{Find $\wt{r} \in \Delta^n$ s.t. $\|\wt{r}-r\|_1 \leq \epsilon_0 / 4$ and $\min_{i\in [n]} \wt{r}_i \geq \epsilon_0 / (8n)$.} 
\State $\wt{r} \gets (1-\frac{\epsilon_0}{8} ) (r+\frac{\epsilon_0}{n (8-\epsilon_0 )}\mathbf{1}_n )$\label{line:find_r_c_sym}

\State $(u,L,w) \gets \textsc{SinkhornAlgorithm}(\wt{r}$,$\epsilon_0 / 2$) \Comment{Algorithm \ref{alg:sinkhorn_treewidth_sym}} \label{line:calculate_b_sym}
\State \Comment{Note that $u,v,L,w$ is an implicit representation of $B$, i.e., $\diag(e^{u}) (L_A L_A^\top) \diag(e^{u})  +\diag(e^{u}) (w w^\top) \diag(e^{u})$}
\State  $(p$, $X, Y, w, u)$ $\gets \textsc{Round}(u,L,w,r)$ \Comment{Algorithm \ref{alg:round_sym}} \label{lin:round_sym}
\State \Return $(p$, $X, Y, L, w, u)$   \Comment{We return $\hat{X}$ in an implicit way, i.e., $\hat{X} := XBY + p p^\top / \|p\|_1$ }
\EndProcedure
\end{algorithmic}
\end{algorithm}

\begin{algorithm}[!ht]
\caption{Rounding of the projection of $B$ on $\mathcal{U}$ for symmetric distribution}\label{alg:round_sym}
\begin{algorithmic}[1]
\Procedure{RoundSym}{$u \in \R^n,  L \in \R^{n \times n} ,w \in \R^{n},r \in \R^n$} \Comment{Lemma \ref{lem:round_sym}   
}
\State \Comment{$L$ is a lower triangular matrix that only has $O(n \tau)$ nonzeros}
\State \Comment{We never explicit write $B$. $B$ can implicitly represented by $\diag(e^{u}) (L_A L_A^\top) \diag(e^{u})  +\diag(e^{u}) (w w^\top) \diag(e^{u})$}
\State $X \gets \diag(x)$ with $x_i = \min\{\frac{r_i}{(B\mathbf{1}_n)_i}, 1\}$
\State $B_0 \gets XB$ \Comment{We only implicitly construct $B_0$}
\State $Y \gets \diag(y)$ with $y_j =  \min\{\frac{r_j}{(B_0^\top \mathbf{1}_n)_j} , 1\}$
\State $B_1 \gets B_0 Y$ \Comment{We only implicitly construct $B_1$}
\State $p \gets r-B_1 \mathbf{1}_n$  
\State \Return $p, X, Y, w, u$ \Comment{We return $G$ in an implicit way, i.e., $G := B_1 + p p^\top / \|p\|_1$ }
\EndProcedure
\end{algorithmic}
\end{algorithm}

Next, we show the running time of the rounding algorithm (Algorithm \ref{alg:round_sym}) for symmetric case.
\begin{lemma}[An improved version of of Lemma 7 in \cite{anr17}]\label{lem:round_sym}
Given $r \in \triangle_n$, $B \in \R_+^{n \times n}$, $u \in \R^{n}$, there is an algorithm (Algorithm \ref{alg:round_sym}) outputs 
\begin{itemize}
\item a diagonal matrix $X \in \R^{n \times n}$
\item a diagonal matrix $Y \in \R^{n \times n}$
\item a lower triangular matrix $L_A$
\item vectors $u, w, p \in \R^n$
\end{itemize}
such that $G \in \mathcal{U}_{r}$  can be constructed (implicitly) by 
\begin{align*}
\hat{X} = X ( \diag(e^u) L_A L_A^\top \diag(e^u) + \diag(e^u) (w w^\top) \diag(e^u) ) Y  + p p^\top / \|p\|_1
\end{align*}
that satisfying 
\begin{align*}
     \|G-B\|_1 \leq 2(\|B \mathbf{1}_n -r\|_1) 
\end{align*}
in $O(n \tau)$ time.
\end{lemma}
\begin{proof}
    Similar to the proof of Lemma \ref{lem:round_correct} and Lemma \ref{lem:round}, here the two distribution are identical, i.e., $c=r$.
\end{proof}

Overall, we provide the running time of the algorithm (Algorithm \ref{alg:ot_dis_sym}) that approximate the OT for symmetric case.
\begin{theorem}\label{thm:runningtime_ot_sym}
There is an algorithm (Algorithm \ref{alg:ot_dis_sym}) takes cost matrix $C = MM^\top = \in \R^{n \times n}$, an $n$-dimensional simplex $r $ as inputs and  outputs 
\begin{itemize}
\item a diagonal matrix $X \in \R^{n \times n}$
\item a diagonal matrix $Y \in \R^{n \times n}$
\item a lower triangular matrix $L_A$
\item vectors $u, w, p \in \R^n$
\end{itemize}
such that $\hat{X} \in \mathcal{U}(r)$  can be constructed (implicitly) by 
\begin{align*}
\hat{X} = X ( \diag(e^u) L_A L_A^\top \diag(e^u) + \diag(e^u) (w w^\top) \diag(e^u) ) Y  +  p p^\top / \|p\|_1
\end{align*}
that satisfying Eq.~\eqref{eq:app_ot} in 
    \begin{align*}
        O( n \tau^2 + \epsilon^{-2} n\tau \|C\|_\infty^2 \ln n  )
    \end{align*}
    time.
\end{theorem}
\begin{remark}
If we don't care about the output format to be lower-triangular matrix, then the additive term $n\tau^2$ can be removed.
\end{remark}

\begin{proof}
By using Theorem \ref{thm:per_sinkhorn_treewidth_sym}, we have the running time of Line \ref{line:calculate_b_sym} is $O(n \tau \cdot T)$, where $T$ is the total number of Sinkhorn's algorithm iterations. By using Lemma \ref{lem:round_sym}, the running time of Line \ref{lin:round_sym} is $O(n \tau)$. The rest of the proof is similar to the proof of Similar to the proof Theorem \ref{thm:runningtime_ot}.
\end{proof}

\ifdefined\isarxiv
\bibliographystyle{alpha}
\bibliography{ref}
\else
\bibliography{ref}

\newcommand{\etalchar}[1]{$^{#1}$}
\begin{thebibliography}{AZLOW17}

\bibitem[ACB17]{acb17}
Martin Arjovsky, Soumith Chintala, and L{\'e}on Bottou.
\newblock Wasserstein generative adversarial networks.
\newblock In {\em International conference on machine learning}, pages
  214--223. PMLR, 2017.

\bibitem[ANWR17]{anr17}
Jason Altschuler, Jonathan Niles-Weed, and Philippe Rigollet.
\newblock Near-linear time approximation algorithms for optimal transport via
  sinkhorn iteration.
\newblock {\em Advances in neural information processing systems}, 30, 2017.

\bibitem[AZLOW17]{alow17}
Zeyuan Allen-Zhu, Yuanzhi Li, Rafael Oliveira, and Avi Wigderson.
\newblock Much faster algorithms for matrix scaling.
\newblock In {\em 2017 IEEE 58th Annual Symposium on Foundations of Computer
  Science (FOCS)}, pages 890--901. IEEE, 2017.

\bibitem[BBR06]{bbr06}
Federico Bassetti, Antonella Bodini, and Eugenio Regazzini.
\newblock On minimum kantorovich distance estimators.
\newblock {\em Statistics \& probability letters}, 76(12):1298--1302, 2006.

\bibitem[BGHK95]{bghk95}
Hans~L Bodlaender, John~R Gilbert, Hj{\'a}lmtyr Hafsteinsson, and Ton Kloks.
\newblock Approximating treewidth, pathwidth, frontsize, and shortest
  elimination tree.
\newblock {\em Journal of Algorithms}, 18(2):238--255, 1995.

\bibitem[BGKL17]{bgkl17}
J{\'e}r{\'e}mie Bigot, Ra{\'u}l Gouet, Thierry Klein, and Alfredo L{\'o}pez.
\newblock Geodesic pca in the wasserstein space by convex pca.
\newblock In {\em Annales de l'Institut Henri Poincar{\'e}, Probabilit{\'e}s et
  Statistiques}, volume~53, pages 1--26. Institut Henri Poincar{\'e}, 2017.

\bibitem[BJKS18]{bjks18}
Jose Blanchet, Arun Jambulapati, Carson Kent, and Aaron Sidford.
\newblock Towards optimal running times for optimal transport.
\newblock {\em arXiv preprint arXiv:1810.07717}, 2018.

\bibitem[Bod94]{b94}
Hans~L Bodlaender.
\newblock A tourist guide through treewidth.
\newblock {\em Acta cybernetica}, 11(1-2):1, 1994.

\bibitem[BPC16]{bpc16}
Nicolas Bonneel, Gabriel Peyr{\'e}, and Marco Cuturi.
\newblock Wasserstein barycentric coordinates: histogram regression using
  optimal transport.
\newblock {\em ACM Trans. Graph.}, 35(4):71--1, 2016.

\bibitem[CMTV17]{cmtv17}
Michael~B Cohen, Aleksander Madry, Dimitris Tsipras, and Adrian Vladu.
\newblock Matrix scaling and balancing via box constrained newton's method and
  interior point methods.
\newblock In {\em 2017 IEEE 58th Annual Symposium on Foundations of Computer
  Science (FOCS)}, pages 902--913. IEEE, 2017.

\bibitem[Cut13]{c13}
Marco Cuturi.
\newblock Sinkhorn distances: Lightspeed computation of optimal transport.
\newblock {\em Advances in neural information processing systems}, 26, 2013.

\bibitem[Dav06]{d06}
Timothy~A Davis.
\newblock {\em Direct methods for sparse linear systems}.
\newblock SIAM, 2006.

\bibitem[DGK18]{dgk18}
Pavel Dvurechensky, Alexander Gasnikov, and Alexey Kroshnin.
\newblock Computational optimal transport: Complexity by accelerated gradient
  descent is better than by sinkhorn’s algorithm.
\newblock In {\em International conference on machine learning (ICML)}, pages
  1367--1376. PMLR, 2018.

\bibitem[DLY21]{dly21}
Sally Dong, Yin~Tat Lee, and Guanghao Ye.
\newblock A nearly-linear time algorithm for linear programs with small
  treewidth: A multiscale representation of robust central path.
\newblock In {\em STOC}, 2021.

\bibitem[ESS17]{ess17}
Johannes Ebert, Vladimir Spokoiny, and Alexandra Suvorikova.
\newblock Construction of non-asymptotic confidence sets in 2-wasserstein
  space.
\newblock {\em arXiv preprint arXiv:1703.03658}, 2017.

\bibitem[FLS{\etalchar{+}}18]{fls+18}
Fedor~V Fomin, Daniel Lokshtanov, Saket Saurabh, Micha{\l} Pilipczuk, and
  Marcin Wrochna.
\newblock Fully polynomial-time parameterized computations for graphs and
  matrices of low treewidth.
\newblock {\em ACM Transactions on Algorithms (TALG)}, 14(3):1--45, 2018.

\bibitem[FZM{\etalchar{+}}15]{fzm+15}
Charlie Frogner, Chiyuan Zhang, Hossein Mobahi, Mauricio Araya, and Tomaso~A
  Poggio.
\newblock Learning with a wasserstein loss.
\newblock {\em Advances in neural information processing systems}, 28, 2015.

\bibitem[GLN94]{gln94}
Alan George, Joseph Liu, and Esmond Ng.
\newblock Computer solution of sparse linear systems.
\newblock {\em Oak Ridge National Laboratory}, 1994.

\bibitem[GPC18]{gpc18}
Aude Genevay, Gabriel Peyr{\'e}, and Marco Cuturi.
\newblock Learning generative models with sinkhorn divergences.
\newblock In {\em International Conference on Artificial Intelligence and
  Statistics}, pages 1608--1617. PMLR, 2018.

\bibitem[GS22]{gs22}
Yuzhou Gu and Zhao Song.
\newblock A faster small treewidth sdp solver.
\newblock {\em arXiv preprint arXiv:2211.06033}, 2022.

\bibitem[Kan42]{j42}
Leonid~V Kantorovich.
\newblock On the translocation of masses.
\newblock In {\em Dokl. Akad. Nauk. USSR (NS)}, volume~37, pages 199--201,
  1942.

\bibitem[KPT{\etalchar{+}}17]{kpt+17}
Soheil Kolouri, Se~Rim Park, Matthew Thorpe, Dejan Slepcev, and Gustavo~K
  Rohde.
\newblock Optimal mass transport: Signal processing and machine-learning
  applications.
\newblock {\em IEEE signal processing magazine}, 34(4):43--59, 2017.

\bibitem[KSKW15]{kskw15}
Matt Kusner, Yu~Sun, Nicholas Kolkin, and Kilian Weinberger.
\newblock From word embeddings to document distances.
\newblock In {\em International conference on machine learning}, pages
  957--966. PMLR, 2015.

\bibitem[LSZ{\etalchar{+}}22]{lszzz22}
Sixue Liu, Zhao Song, Hengjie Zhang, Lichen Zhang, and Tianyi Zhou.
\newblock Space-efficient interior point method, with applications to linear
  programming and maximum weight bipartite matching.
\newblock arXiv preprint arXiv:2009.06106, 2022.

\bibitem[MMC16]{mmkc16}
Gr{\'e}goire Montavon, Klaus-Robert M{\"u}ller, and Marco Cuturi.
\newblock Wasserstein training of restricted boltzmann machines.
\newblock {\em Advances in Neural Information Processing Systems}, 29, 2016.

\bibitem[PW09]{pw09}
Ofir Pele and Michael Werman.
\newblock Fast and robust earth mover's distances.
\newblock In {\em 2009 IEEE 12th international conference on computer vision},
  pages 460--467. IEEE, 2009.

\bibitem[PZ16]{pz16}
Victor~M Panaretos and Yoav Zemel.
\newblock Amplitude and phase variation of point processes.
\newblock {\em The Annals of Statistics}, 44(2):771--812, 2016.

\bibitem[Qua18]{q18}
Kent Quanrud.
\newblock Approximating optimal transport with linear programs.
\newblock {\em arXiv preprint arXiv:1810.05957}, 2018.

\bibitem[RS10]{rs10}
Neil Robertson and Paul Seymour.
\newblock Graph minors xxiii. nash-williams' immersion conjecture.
\newblock {\em Journal of Combinatorial Theory, Series B}, 100(2):181--205,
  2010.

\bibitem[RTG00]{rtg00}
Yossi Rubner, Carlo Tomasi, and Leonidas~J Guibas.
\newblock The earth mover's distance as a metric for image retrieval.
\newblock {\em International journal of computer vision}, 40(2):99--121, 2000.

\bibitem[SBRL18]{sbrl18}
Maziar Sanjabi, Jimmy Ba, Meisam Razaviyayn, and Jason~D Lee.
\newblock On the convergence and robustness of training gans with regularized
  optimal transport.
\newblock {\em Advances in Neural Information Processing Systems}, 31, 2018.

\bibitem[SRGB14]{srgb14}
Justin Solomon, Raif Rustamov, Leonidas Guibas, and Adrian Butscher.
\newblock Wasserstein propagation for semi-supervised learning.
\newblock In {\em International Conference on Machine Learning}, pages
  306--314. PMLR, 2014.

\bibitem[SYYZ22]{syyz22}
Zhao Song, Xin Yang, Yuanyuan Yang, and Tianyi Zhou.
\newblock Faster algorithm for structured john ellipsoid computation.
\newblock {\em arXiv preprint arXiv:2211.14407}, 2022.

\bibitem[SZRM18]{szrm18}
Tim Salimans, Han Zhang, Alec Radford, and Dimitris Metaxas.
\newblock Improving gans using optimal transport.
\newblock {\em arXiv preprint arXiv:1803.05573}, 2018.

\bibitem[Vil09]{v09}
C{\'e}dric Villani.
\newblock {\em Optimal transport: old and new}, volume 338.
\newblock Springer, 2009.

\bibitem[WPR85]{wpr85}
Michael Werman, Shmuel Peleg, and Azriel Rosenfeld.
\newblock A distance metric for multidimensional histograms.
\newblock {\em Computer Vision, Graphics, and Image Processing},
  32(3):328--336, 1985.

\end{thebibliography}
\bibliographystyle{alpha}

\fi

\newpage
\onecolumn
\appendix




\end{document}